\documentclass[epj,twocolumn]{webofc}
\usepackage[varg]{txfonts}
\woctitle{Hadron Collider Physics symposium 2012}

\def\AFB{${\rm A}_{\rm FB}~$}

\begin{document}
\thanks{Preprint number: CERN-PH-TH/2013-042}

\title{Recent theoretical progress in top quark pair production at hadron colliders}
\author{Alexander Mitov\inst{1}\fnsep\thanks{\email{alexander.mitov@cern.ch}}}
\institute{Theory Division, CERN, CH-1211 Geneva 23, Switzerland}

\abstract{This is a writeup of a plenary talk given at the conference {\it HCP 2012} held November 2012 in Kyoto, Japan. This writeup reviews recent theoretical developments in the following areas of top quark physics at hadron colliders: (a) the forward-backward asymmetry anomaly at the Tevatron, (b) precision top mass determination, (c) state of the art NLO calculations and (d) progress in NNLO calculations.}

\maketitle

\section{Introduction}

Driven by the vast number of top quark pairs produced at the LHC, top physics is entering into a high precision phase. This allows, for the first time, to study top quarks in finest detail. Notably, the experimental advances are matched by theory: new generation of precision calculations have appeared and are beginning to make their way into the experimental analyses and searches for new physics. We review them in the following.

\section{Forward-Backward asymmetry at the Tevatron}

The top quark forward-backward asymmetry \AFB has been measured by the CDF \cite{Aaltonen:2012it} and D0 \cite{Abazov:2011rq} collaborations and was found to differ by more than $2\sigma$ from SM-based theoretical predictions. Many bSM explanations have been proposed. They are reviewed, for example, in Ref.~\cite{Kamenik:2011wt}. In the following we review the status of the various Standard Model (SM) contributions. 

The largest known contribution to \AFB is due to ${\cal O}(\alpha_S^3)$ QCD corrections \cite{Kuhn:1998kw}. As it turns out, they are insufficient for describing the Tevatron data. Electroweak corrections have been revisited \cite{Hollik:2011ps} and found to be rather large, around 20\% of the QCD ${\cal O}(\alpha_S^3)$ correction. Calculations combining strong and electroweak corrections have appeared \cite{Bernreuther:2012sx}. The forward-backward asymmetry has also been studied in related lepton-based observables \cite{Bernreuther:2012sx,Falkowski:2012cu} as well as from a collider-independent prospective \cite{AguilarSaavedra:2012va}. Higher order QCD corrections could also play a role but none have been computed at present (though they are anticipated in the very near future). 

The only information about unknown, beyond ${\cal O}(\alpha_S^3)$ corrections to \AFB comes from soft-gluon resummation. It was shown in \cite{Almeida:2008ug,Ahrens:2011uf} that soft gluon resummation, when matched to leading order top pair production (which itself has zero asymmetry), generates \AFB  that is numerically as large as the full ${\cal O}(\alpha_S^3)$ result.
\footnote{Electroweak Sudakov logarithms have also been resummed \cite{Manohar:2012rs}.}
Moreover, the predicted asymmetry does not seem to depend on the logarithmic accuracy of the resummation. In this context it is very interesting to recall the recent work \cite{Skands:2012mm} where it was shown that non-zero \AFB can be generated purely from the parton shower, in fact, already after the first emission. This is a very interesting finding that surely will add to the debate about the relationship between usual (inclusive) resummations and parton showers. 

The fact that soft corrections contribute very little to \AFB beyond ${\cal O}(\alpha_S^3)$ could be interpreted \cite{Almeida:2008ug,Melnikov:2010iu} as the near absence of higher order corrections to \AFB. This is an interesting possibility that a future dedicated calculation of the ${\cal O}(\alpha_S^4)$ corrections to \AFB should be able to clarify.
  
Various other possibilities that could address \AFB have been considered in the literature. Among them are the so-called BLM corrections \cite{Brodsky:1982gc}. It was claimed in Ref.~\cite{Brodsky:2012ik} that by selecting the scales following this principle, one arrives at a prediction for \AFB which is compatible with the Tevatron measurements. While the BLM approximation is known to be an excellent one in other contexts, its applicability in top physics at hadron colliders has not yet been broadly established.

Another mechanism studied in the literature is the idea that \AFB could be generated from final state interactions, i.e. interactions of the $t\bar t$ pair with the beam remnants \cite{Rosner:2012pi,Mitov:2012gt}. An all-order proof of the cancellation of such interactions was given in \cite{Mitov:2012gt} where it was also shown that the effect of final state interactions on inclusive \AFB is negligible (but could be substantial in cases of strong jet vetoes, for example).

\section{Top mass determination}

Like all fundamental parameters of the SM, the top mass requires careful study and precise determination. Current measurements \cite{Aaltonen:2012ra} put its value at $173.18 \pm 0.94$ GeV which has a fantastic, almost 0.5\%, precision. Given the extensive use of (often leading order) Monte Carlo generators in many of the extractions of $m_{\rm top}$, the question about the precise relationship between the accurately measured $m_{\rm top}$, and a well-defined theoretically top mass emerges. 

The main motivation for improved precision in the determination of $m_{\rm top}$ arises, {\it at present}, not directly in collider physics but in more formal aspects of the SM.
\footnote{The dominant uncertainty in EW precision fits at present is due to $M_W$ \cite{Baak:2012kk}.}
It was demonstrated in Ref.~\cite{Degrassi:2012ry} that the top quark mass, together with $M_{higgs}$ and $\alpha_S$, are the main contributors of uncertainty in predicting the scale of SM vacuum stability's breakdown. This is a particularly interesting question given that a relatively small ${\cal O}(1~ {\rm GeV})$ change in the top quark mass could, in essence, push the validity of the SM by orders of magnitude and up to GUT-level scales. Strong dependence on $m_{\rm top}$ appears also in the so-called Higgs-inflation model \cite{Bezrukov:2007ep,DeSimone:2008ei}. 

A number of approaches for the determination of $m_{\rm top}$ have been discussed in the literature and we briefly review their status in the following.

The most popular approach for determination of the top quark mass is to exploit the class of matrix element methods. Their main drawback, at present, is that the theoretical input used in the construction of the likelihood functions is derived from LO QCD. The effect of missing higher order corrections has not been addressed so far. Applications at NLO QCD are being developed \cite{Campbell:2012cz}.

Another approach is based on the extraction of $m_{\rm top}$ from differential distributions in top quark events. Ideally, such kinematic distributions are strongly dependent on the value of $m_{\rm top}$, are clean experimentally and could be modeled well theoretically. One such method is known as the {\it method with $J/\Psi$ final states} \cite{Kharchilava:1999yj}. This method utilizes the invariant mass of three leptons: one from the $W$ decay and two muons from the decay of the $J/\Psi$ particle resulting (with very low probability $\sim 10^{-5}$) from the fragmentation of the $b$-quark in the decay of the same top quark. Experimentally such a signal is very clean. Its only drawback is the large required number of top events. Theoretically, the distribution can be modeled very cleanly and is now known in NLO QCD \cite{Biswas:2010sa} including finite top-width effects \cite{Denner:2012yc} (the latter affect strongly the tail of the invariant mass distribution but are not very important in the peak region which is most relevant for this method).

Another possibility is to consider lepton distributions in dilepton top pair events since, similarly to the $J/\Psi$ method, they are little sensitive to modeling of hadronic radiation, event reconstruction and shower effects and can be well described with the help of existing complete NLO calculations \cite{Bevilacqua:2010qb,Denner:2012yc}.

A third approach is to extract the top mass from the measured total inclusive cross-section. Such an observable is known to even higher orders and is likely rather insensitive to modeling of hadronic radiation and top width effects. The method has its limit, however, which is dictated by the not-so-strong dependence of the total cross-section with respect to $m_{\rm top}$ (which is around 3\% and is thus not competitive in precision with the methods described above). A number of papers have presented such extractions \cite{Ahrens:2011px,Beneke:2011mq,Langenfeld:2009wd,Beneke:2012wb}.

Finally, it is worth noting that currently we are witnessing a renewed interest and activity towards a better understanding of top mass determination at hadron colliders.

\section{Fully differential NLO calculations}

Due to their extremely short lifetime, top quarks are never measured directly at colliders. Since $V_{tb}\approx 1$, top quarks decay to a $b$ quark and a $W$ boson. Furthermore the $W$ boson decays either hadronically or semileptonically, while the $b$ quark is observed either as part of a jet (that could or could not be tagged as a $b$-jet) or possibly as an isolated $b$-flavored hadron (typically a $B$-meson). Clearly, even in the absence of higher order corrections in the form of additional hadronic radiation, a top pair event has an extremely complicated experimental signature: either 6 jets or 4(2) jets + 1(2) leptons + missing energy. As the large NLO QCD corrections to the $t\bar t$ cross-section suggest, most of the time additional hadronic radiation is emitted either in the top-production or top-decay stage further complicating the structure of the observed final state. It is for this reason that calculations involving $t\bar t$ final states are not realistic and additional tools are needed to relate the observed jets and/or leptons to the assumed top quarks in the event. 

Advances in one-loop calculations made possible NLO QCD corrections of fully differential, associated $t\bar t$ production in some quite complicated processes like $t\bar t +H$ \cite{Beenakker:2001rj,Dawson:2003zu}, $t\bar t+j$ \cite{Dittmaier:2008uj}, $t\bar t+2j$ \cite{Bevilacqua:2010ve}, $t\bar t b\bar b$ \cite{Bevilacqua:2009zn,Bredenstein:2010rs}. 

Eventually, more realistic calculations of fully differential $t\bar t$ production matched, in the narrow width approximation, to top quark decays \cite{Bernreuther:2004jv,Melnikov:2009dn,Bernreuther:2010ny,Campbell:2012uf} appeared. Such calculations have now been done not only for $t\bar t$ but also for several associated production reactions like $t\bar t+j$ \cite{Melnikov:2011qx} and $t\bar t+\gamma$ \cite{Melnikov:2011ta}. Finally, the complete off-shell effects in $t\bar t$ production were computed in \cite{Bevilacqua:2010qb,Denner:2012yc}.

Separately, the NNLO corrections to fully differential top quark decays have also been computed \cite{Gao:2012ja,Brucherseifer:2013iv}, opening the door for future fully differential NNLO calculations of top pair production and decay.

In the last few years progress in the calculation of NLO corrections to top quark processes including parton shower was made \cite{Kardos:2011qa,Garzelli:2011vp,Garzelli:2012bn,Alioli:2011as}. In particular the {\it aMC@NLO} package which is capable of dealing with very complex final states at NLO accuracy, including showering, became publicly available \cite{amcnlo}.

\section{High-precision NNLO results}

Improved experimental precision at the LHC, combined with the available very large number of top pair events, has lead to measurements that have total uncertainty as low as few per-cent. Such precision cannot be matched by NLO calculations and requires NNLO level of precision. 

Significant progress in this direction was made in the last few years. First steps were made through the so-called Coulomb--threshold approximation, where the full NNLO result is approximated by the terms enhanced as $\sim 1/\beta^k$ or $\sim \ln^n(\beta)$ in the limit of small relative velocity $\beta$. These terms are predicted from soft gluon resummation with NNLL accuracy \cite{Beneke:2009rj,Czakon:2009zw,Ahrens:2010zv}. Correct expansion through NNLO was first derived in \cite{Beneke:2009ye}. A number of phenomenological studies appeared \cite{Ahrens:2010zv,Langenfeld:2009wd,Beneke:2010fm,Kidonakis:2010dk,Ahrens:2011mw,Ahrens:2011px,Kidonakis:2011ca,Beneke:2011mq,Cacciari:2011hy}. Overall, a significant spread due to differences in subleading terms can be observed between the various predictions.

Resummed calculations can be performed with the help of the programs {\tt Top++} \cite{Czakon:2011xx} and {\tt TOPIXS} \cite{Beneke:2012wb} and pure fixed order ones also with the program {\tt Hathor} \cite{Aliev:2010zk}.

Another NNLO approximation is based on the high-energy limit of the cross-section, where the partonic energy $s$ is much larger than the mass of the top quark, i.e.  $\rho\equiv 4m^2/s\approx 0$. It is well known \cite{Nason:1987xz,Catani:1990xk,Collins:1991ty,Catani:1990eg,Catani:1993ww,Catani:1994sq} that in this high-energy limit the cross-section develops logarithmic singularity that can be independently predicted: 
\begin{equation}
\sigma_{\rm tot} \approx c_1\ln(\rho) + c_0 + {\cal O}(\rho) \, .
\label{HElimit}
\end{equation}
The analytical result for the constant $c_1$ in all partonic channels has been given in Ref.~\cite{Ball:2001pq}.

In Ref.~\cite{Moch:2012mk} approximate numerical values for the constant $c_0$ in all partonic channels were derived. Subsequently, the constants for the $q\bar q$ and $qg$ channels  were computed  also numerically in Refs.~\cite{Czakon:2012zr,Czakon:2012pz} and found to agree with the prediction of Ref.~\cite{Moch:2012mk}, within the numerical uncertainties. Phenomenology based on this approximation is a very delicate issue since for top pair production at LHC energies the partonic fluxes vanish in the region where the high-energy approximation applies. See Refs.~\cite{CMS:alpha_s,Czakon:2012pz,Czakon:2013vfa} for more details.

The exact NNLO result for top pair production at the Tevatron was derived in Ref.~\cite{Baernreuther:2012ws}. This calculation reduced the scale uncertainty down to about $\sim \pm 2.7 \%$. A good agreement between the SM prediction and Tevatron data was observed there. Subsequently the ${\cal O}(\alpha_S^4)$ corrections due to the all-fermionic scattering channels was computed \cite{Czakon:2012zr} as well as the $qg$-initiated one \cite{Czakon:2012pz}. Despite the very large size of the partonic flux for the all-fermionic channels their contributions to the cross-section were found to be at the per-mill level. The contribution of the $qg$-scattering channel is at the $1\%$ level. Interestingly, its absolute value is as large - or even larger - than the lower-order contribution from the same process.

Clearly, for complete NNLO phenomenology at the LHC, the $gg$-initiated partonic reaction is needed since it dominates due to the size of the gluon flux at the LHC. The results for this reaction are expected very soon.

Finally, we would like to stress that the NNLO calculations \cite{Baernreuther:2012ws,Czakon:2012zr,Czakon:2012pz} for the total inclusive cross-section can be directly extended to arbitrary differential distributions of the top pair. Down the road, by combining such results with the already available fully differential top quark decay \cite{Gao:2012ja,Brucherseifer:2013iv}, it would be very natural to construct NNLO partonic Monte Carlo generator for realistic final states in the narrow width approximation.

\section{Outlook}

The theoretical understanding of top physics at hadron colliders improved substantially during the last few years. We are witnessing a renewed interest in the problem of the precise determination of the top quark mass: theoretical progress, coupled with the number of precise measurements already available from the LHC, suggests new developments are plausible and could be anticipated soon. Outstanding questions in precision top physics are also being resolved: a comparison of the available experimental measurements of the total inclusive cross-section with the complete NNLO QCD calculation suggests analyses at full data set - at both 7 and 8 TeV as well as their ratio \cite{Mangano:2012mh} - would be of particular value. A presumably even more precise future Atlas/CMS cross-section combination could be very helpful in a number of precision SM analyses and searches for new physics. 

In the next couple of years, a fully differential description of top pair production in NNLO QCD is plausible, and it will further refine the currently available fully differential NLO predictions. In particular the calculation of the dominant uncertainty to \AFB (expected to appear in the near future) should help clarify this outstanding discrepancy between SM predictions and measurements from the Tevatron.

\bigskip
The author would like to thank the organizers for their invitation to this exciting conference in beautiful Kyoto and for partial support. The work of A.M. is supported by ERC grant 291377 ``LHCtheory: Theoretical predictions and analyses of LHC physics: advancing the precision frontier".

\end{document}